**Purposeful remixing with generative AI:**

**Constructing designer voice in multimodal composing**

Xiao Tan, Duke University, xiao.tan@duke.com

Wei Xu, The University of Arizona, weixu1@arizona.edu

Chaoran Wang, Colby College, chaowang@colby.edu

**Abstract**

Voice, the discursive construction of the writer's identity, has been extensively studied and theorized in composition studies. In multimodal writing, students are able to mobilize both linguistic and non-linguistic resources to express their real or imagined identities. But at the same time, when students are limited to choose from available online resources, their voices might be compromised due to the incompatibility between their authorial intentions and the existing materials. This study, therefore, investigates whether the use of generative AI tools could help student authors construct a more consistent voice in multimodal writing. In this study, we have designed a photo essay assignment where students recount a story in the form of photo essays and prompt AI image-generating tools to create photos for their storytelling. Drawing on interview data, written reflection, written annotation, and multimodal products from seven focal participants, we have identified two remixing practices–*layering* and *blending*–through which students attempted to establish a coherent and unique voice in writing. The study sheds light on the intentional and discursive nature of multimodal writing with AI as afforded by the technological flexibility, while also highlighting the practical and ethical challenges that could be attributed to students' insufficient prompt and multimodal literacy and the innate limitations of AI systems. This study provides important implications for incorporating AI tools in designing multimodal writing tasks.

**1. Introduction**

Voice, the discursive construction of the writer's identity, is an important element in writing communication (Matsuda, 2015). Having an appropriate voice is widely considered a desirable feature in advanced and mature writing (e.g., Morton & Storch, 2019). As a result, it is not uncommon for writing teachers to teach and assess, whether explicitly or otherwise, voice in students' text-based writing (Matsuda & Jeffery, 2012). With the added benefits of using non-textual resources in multimodal writing, students could enjoy greater flexibility in developing a wider range of identities in their multimodal works (Kim & Li, 2021; Author, year). But at the same time, scholars have also noted that the construction of a coherent voice might be constrained by the common rexming practices in multimodal writing (Hafner, 2015). In other words, students' voices could be compromised when they have to choose from online materials that are designed for other rhetorical situations (Hafner, 2015).

The rapid development of generative artificial intelligence (GAI) technology, with its potentials of instantly generating human-like texts and images based on natural language prompt (Ranade & Eyman, 2024), provides a possible way out of the dilemma: instead of constraining themselves to existing resources, students could use AI technology to generate visual materials that fit more appropriately into a specific rhetorical situation and fulfill their authorial intentions. Moreover, in order to produce images that best align with their rhetorical intentions, students have to engage in a recursive process of feeding prompts, analyzing and selecting images, and revising prompts (Kang & Yi, 2023). The decisions made during this process could arguably result in a more coherent voice in the final products. Based on these hypotheses, we designed this study to investigate whether using AI image-generating tools could contribute positively to the construction of voice in multimodal writing. The next section reviews the theoretical discussions and empirical findings of voice in written and multimodal communication.

## 2. Literature Review

### 2.1 Voice in Written and Multimodal Communication

Voice in written discourses has been defined and theorized in multiple ways following different theoretical orientations (Matsuda, 2015; Tardy, 2012). The individual aspect of voice embraces the idea of voice as an individual property in writing, which is rooted in the analogy of the metaphorical written voice and the literal human voice as a distinct personal trait. In this sense, the concept of voice is therefore associated with qualities such as *authenticity* and *sincerity* (Elbow, 2007). The social perspective of voice, on the other hand, considers the voice of a written text as grounded in the discourse community. Closely related to the socially constructed voice is Ivanic's (1998) notion of *discoursal self*, or "the impression–often multiple, sometimes contradictory–which [writers] consciously or unconsciously convey of themselves in a particular written text" (p. 25). In other words, written voice is constructed through the writers' deployment of "community-sensitive resources to represent themselves, their positions, and their readers" (Hyland, 2008, p. 20). The social aspect of voice also acknowledges the multiplicity of voice within a single text, "as writers resort to and blend many voices in deliberative or unconscious ways as they write" (Tardy, 2012, p. 39). Bakhtin's well-known notion of *heteroglossia* underlines the coexistence of multiple voices, as informed by the inherent ideologies of our social, cultural, and political backgrounds, within the same utterance (Park-Fuller, 1986).

The third theoretical orientation–the dialogical perspective–sees voice as neither owned by the individual writer nor social conventions; it is rather constructed out of the interaction between these two. Matsuda (2001) defines voice as "the amalgamation effect of the use of discursive and non-discursive features that language users choose, deliberately or otherwise, from socially available yet ever-changing repertoires" (p. 41), highlighting the role that the reader could play in constructing the writer's voice. From the dialogic perspective, while the author might have the liberty to choose from various discursive and non-discursive choices, the construction of their voice is ultimately rendered by the reader (Matsuda, 2015).

Voice in multimodal writing has attracted some, albeit limited, attention in the field of composition studies. While scholars seem to agree that the use of multimodal features, such as images, video, and sound effects, could also contribute to the construction of voice (Hafner, 2015; Matsuda, 2015), empirical studies on this topic are scant. In classroom-based research on multimodal writing, several scholars have noted that multimodal writing assignments could afford great opportunities for students to exercise their autonomy and mobilize a wide range of semiotic resources, thus contributing to the development of writer's voice and identity (Author, year; Bloch, 2018, 2021; Kim & Li, 2021; Nelson, 2008).

Very few studies have attempted to define voice in multimodal context and analyze this concept systematically. Inquiring into "academic voice" in the multimodal writing produced by college students, Archer (2013) likens voice to Kress' notion of "authorial design," describing it as "the process of giving shape to the interests, purposes, and intentions of the rhetor in relation to the semiotic resources available for realizing/materializing these purposes as apt material, complex signs, texts for the assumed characteristics of a specific audience" (p. 161). She further discussed how academic voice could be manifested in authorial engagement, modality, and citation. The first aspect, authorial engagement, could be realized by using certain attitude markers and personal pronouns, as well as choosing image layout that invites the reader's participation (Archer, 2013). Modality, which indicates "degrees of certainty and expression of obligation" (p. 154), is actualized through the use of hedging phrases and images of different truth values. Lastly, citation in both verbal and visual modes involves appropriating the original sources into one's own argument to make new meaning (Archer, 2013). Although providing new insights into the concept, Archer's discussion of voice in multimodal communication relies only on the textual analysis of exemplary works and does not consider the agentive role that students play when constructing their own voices.

## 2.2 Remixing and AI-Assisted Composing

In digital writing scholarship, *remix* encompasses the practices of appropriating and reworking existing cultural materials to create new works (Edwards, 2016; Hafner, 2015). While the act of remixing has existed throughout the history of writing, the widespread use of digital technologies has greatly increased the ease with which one can alter and repurpose existing resources, to a point where several scholars have argued that remixing is the new norm of writing in the digital age (Edwards, 2016; Palmeri, 2012). Given the ubiquitousness of remixing, scholars have proposed frameworks to better comprehend what is involved in this practice (Edwards, 2016; Hafner, 2015). Based on the analysis of student-created videos in an English for Science course, Hafner's (2015) envisioned the remix practices to include four interconnected types of practice: mixing sources (i.e., chunking), mixing modes (i.e., layering), mixing genres (i.e., blending), and mixing cultural resources (i.e., intercultural blending). Another well-cited framework, proposed by Edwards (2016), theorizes remixing into four categories, namely assemblage, reappropriation, redistribution, and genre play. Both authors have advocated for more pedagogical attention to remixing in writing classrooms using the frameworks as a starting point (Edwards, 2016; Hafner, 2015). Writing teachers have also shared various activities that require students to modify certain parts of the existing artifacts to create new meanings and derive new rhetorical significance (Dusenberry et al., 2015; Palmeri, 2012; Shaw, 2022).

While remixing in writing assignments has opened doors to creative expressions and engaging learning experiences, it is not without problems (Hafner, 2015; Nelson, 2008). For example, Hafner (2015) noticed a lack of coherence in the videos created by his English as a Second Language (ESL) students, which he attributed to the fact that students had to select from repertoires of existing resources and materials that may not be coherent with the rest of the video. As a result, students' voices might be "overpowered by that of the original material," and "[t]he resulting text may seem more like a patchwork jumble than a coherent whole" (Hafner, 2015, p. 504). In Nelson's study (2008), a focal student named Jirou reported that a chosen image was inadequate in supporting his main point in storytelling. He then made small yet important changes to the image to better align it with the linguistic message. While Jirou's case could be seen as a successful example of manipulating existing materials, Nelson (2008) also took into account a less successful case where the relation between different modes is unclear, concluding that "multimedia communication can be both a hindrance to and a facilitator of the expression of authorial voice (understood metaphorically) as well as linguistic expression by implication" (pp. 78-79).

Acknowledging the limitations of remixing with already-existing resources, we propose to use the AI-powered image generating tool as an alternative to achieve a more coherent voice compatible to students' rhetorical intentions. Recent advancements in multimodal generative AI models (e.g., Mid-journey, DALLE-3) allow for the creation of visual images through linguistic inputs. For instance, a writer can craft a detailed prompt, outlining the intended visual content such as a specific style, colors, characters, setup, background, and lighting. Upon receiving the prompt, the AI then generates a visual representation that aligns with the described specifications. Such functionality of text-to-image AI tools has great potentials for writers to visualize ideas in ways that orchestrate their written texts, effectively conveying their meanings in multimodal approaches. Furthermore, the low-floor and high-ceiling principles of these text-to-image generation AI tools allow users without professional digital design training to easily engage with the technology while being able to reach sophisticated outcomes (Vartiainen & Tedre, 2023). However, as noted by Kang and Yi (2023), fine-tuned prompt literacy, i.e., "the trained ability or knowledge to appropriately and effectively formulate and adjust prompts" is needed to generate desired outputs that best suits the writers' specific purposes (p. 2). The process of fine-tuning a prompt, similar to writing a written text, is also iterative as a writer needs to refine their prompts to get the image outputs most aligned with their text and their rhetorical purposes.

### 2.3 The Current Study

The current study explores how AI tools could be leveraged to teach the concept of voice in multimodal writing. To this end, we have designed a writing task that requires students to produce a photo essay with the assistance of Midjourney, the details of which will be provided in the next section. In this study, we adopt Archer's (2013) definition of voice in multimodal texts as "individual agency operating within contextual and semiotic possibilities and constraints" (p. 161). Archer's definition aligns with the social constructivist perspective of voice that "is concerned with how the social conventions are appropriated by individual writers as they respond to the particular rhetorical situation in the process of writing" (Matsuda, 2015, p. 149). While we acknowledge that voice is a dynamic and complex construct encompassing multiple interrelated aspects, we believe that it is useful to examine each of them separately in a single study (Matsuda, 2015). In this study, our primary focus is on the writer's choices in negotiating voices when creating multimodal texts, although the intended voice might be at odds with what the readers perceive. More specifically, this study seeks to answer the following questions:

1) How do students construct their voices in multimodal composing with the assistance of AI-powered image generating tools?

2) What do students perceive as the affordances of using AI-powered image generating tools in multimodal composing?

3) What challenges do students encounter while using AI-powered image generating tools to express voice?

## 3. Method

### 3.1 Research Setting

The photo essay assignment was implemented by the first author in two sections of her first-year composition (FYC) course at a private university in the Eastern United States. The two sections, themed around "language and identity," consist of three major projects that engage students in the critical discussion, analysis, and investigation of language use. The second project, in which our focal assignment was included, invites students to explore how the construction of identity is shaped by the use of language variations. In the fall semester of 2023 when the study was conducted, the first author had a small group of 21 students across the two sections coming from diverse demographic, cultural, and linguistic backgrounds.

### 3.2 The Photo Essay Assignment and Teaching Sequence

The photo essay assignment asked students to choose one of the three academic publications[1] that used narrative inquiry and adapt the story into a photo essay with the assistance

---

[1] The three articles are (1) "When Queer Meets Teacher: A Narrative Inquiry of the Lived Experience of a Teacher of English as a Foreign Language" by Lin et al. (2) "'If You Don't Have English, You're Just as Good as a Dead

of AI image-generating tools. This assignment took a total of four 75-minute class sessions in two weeks. In the first class, the instructor introduced visual rhetoric and the genre of photo essay, leading to students analyzing four images and a photo essay from *The Chronicle of Higher Education* using genre analysis. The second class focused on the discussion of voice based on Tardy's "Current Conceptions of Voice." The third class introduced Midjourney and image generation techniques for photo essays. In the final class, students showcased their photo essays in an exhibition format, offering feedback to each other and reflecting on their choices and definitions of voice in multimodal writing on a shared document. More detailed descriptions of this assignment could be found in Author (year).

### 3.3 Data Collection and Analysis

Data of this study consist primarily of semi-structured interviews with seven focal participants. The data is complemented by the participants' photo essays, their written reflections, their annotations on their own works, and a teaching journal kept by the first author. Seven students (Table 1) volunteered to take an online interview with one of the co-authors. The semi-structured interviews aim to solicit more information about students' writing processes, decision making, as well as their experiences of using AI tools for this particular task (see Appendix I for interview protocol). All interviews were audio recorded with students' consent, which yielded a total length of 214 minutes of recordings.

Students annotated on their photo essays during the last class session. The prompt they received for making annotations was to justify the decisions they made regarding the choices of visuals and written texts. On a shared Google document, students reflected on their writing experiences and theorized their understanding of voice. Finally, in the teaching journal, the first

---

Person': A Narrative of Adult English Language Literacy within Post-Apartheid Africa" by Kaiper. (3) "Narrative and Identity in the 'Language Learning Project'" by Coffey and Street.

author documented her observations of students' engagement and reaction and whenever necessary formed speculations regarding the effectiveness and/or ineffectiveness of teaching.

*Table 1*

Interviewees' demographic information

| Name | Gender | Photo Essay topic |
|------|--------|-------------------|
| Ana | F | Thuli's journey of learning English in the post-apartheid Africa |
| Nathen | M | Sue's journey of becoming a language teacher |
| Elena | F | Jack's journey of finding his queer and professional identity |
| Linda | F | Paul's experience as being an expatriate in Germany |
| Emma | F | Jack's journey of finding his queer and professional identity |
| Harry | M | Thuli's journey of learning English in the post-apartheid Africa |
| Sandy | F | Jack's journey of finding his queer and professional identity |

Data analysis started with transcribing the interview recordings verbatim. To answer the first research question, interview transcripts were coded deductively by identifying the remixing practices in Hafner's model (2015). The coding of the second and third questions was conducted inductively using thematic analysis approach (Braun & Clarke, 2006). After the initial round of coding, we performed focused coding to further refine the categories identified in the first round. Moreover, the findings identified in students' interview data were triangulated by other data sources to strengthen research rigor. All three authors were involved in the coding process, and any disagreement in coding was resolved before the findings were finalized. To ensure that students' input was interpreted accurately, member checks were conducted with each participant.

**4. Findings**

***4.1 Students Constructing Voices in Photo Essays***

Students have demonstrated complex understandings about the key concept of voice. The analysis of students' reflections reveals nuances in how voice is theorized and executed in multimodal writing. Three students (Elena, Sandy, and Emma) place an emphasis on author's intent and persona, with Elena beautifully articulating that "the only way a reader can understand what the author's voice is, is through recognition of the author's interpretation of certain ideas and the intent with which they presented in a medium other than words" (Elena, reflection). Nathan and Linda highlight the textual aspect of voice. Nathan, for example, believes that discussing "less-commonly known perspectives" and "using alternative formatting features" could contribute to developing a unique voice. Lastly, Ana and Harry took into consideration the role of the audience in voice construction. Ana defined voice as "the unique way the author engages with the audience through a piece of writing," while Harry envisions his work as "providing an interactive, engaging experience for the readers that would be otherwise impossible to elicit with voice in non-multimodal communication" (reflection). Informed by the understanding of voice in multimodal writing, students employed two types of remix to purposefully construct their voices in photo essays: *layering* and *blending*.

**Layering**. Layering, also known as mixing modes, involves "the appropriation of visuals from one source and combining them with a student-generated narration" (Hafner, 2015, p. 503). When remixing modes, students took into consideration the relationships between text and images, explaining broadly that text and images should "work together" (Ana, Harry, Linda, and Emma). More specifically, students described the relationships between these two as *complementary*, envisioning that words and images could mutually reinforce each other (Ana, Emma, Elena, Harry, and Nathen). Some students paid special attention to maintaining a consistent tone across text and images (Elena, Emma, Harry, and Sandy). For example, Elena explained in the interview: "So photos, in that sense, kind of match the tone of that sentiment, the struggle, the conflict, the pain being hidden. And that's why they [text and images] are matching together. That's the relationship." Elena's understanding is well reflected in her photo essay. Figure 1 shows an excerpt of her work, in which the main character Jack is portrayed as being in a closet alone. The accompanying text gives an overview of Jack's background, highlighting the struggles that he had been through in gaining recognition for his queer identity.

Figure 1

*An excerpt from Elena's photo essay*

Jack was brought up in a small, close-knit village as a shy youngster. At twenty, he had a long-term heterosexual relationship before moving to South Korea to further his career as a teacher and began to experience instances that made him question his sexuality. Finally, he came to terms with his sexual orientation--but remained closeted until moving to Bangkok at 28, when he came out to his parents. Though they had their misgivings, they were mostly supportive of it. However, his identity was still mostly obscured from his parents since Jack "never had that habit, not now either, that I go into details about my private life." Jack then went to America for two years to get a master's degree in TESOL and finally decided to come back to settle in Thailand, because, "one reason is that Thailand is comfortable for gays."

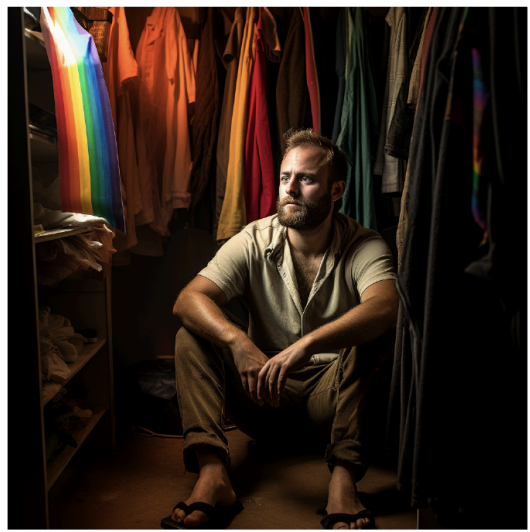

A closer look at the intended design and functions of images reveals students' sophisticated decision making in using AI technologies in multimodal writing. To start with, students approached generating images with a clear idea of the output, attending especially to the overall *tone* of images. As students explained in the interviews, they wanted their images to look realistic (Sandy and Nathen), positive and bright (Harry), historical (Ana), fantasy-like (Emma), dark and gloomy (Linda and Elena). When prompting AI systems to generate images, some students purposefully added words and phrases to achieve their desired outcomes. For example,

Nathen included the word "realistic" to make his images photo-like; Ana, on the other hand, "focus(ed) more toward something that looks more historical, looks more real, simply because the story was that way" (interview). It is interesting to note that even when remediating the same story, students adopted drastically different approaches. For example, the works of Emma, Elena, and Sandy, while all based on the story of Jack, seem to foreground different aspects and therefore convey different messages through the use of visuals. Emma's work places an emphasis on the protagonist's struggles and suffering as a member of the minority group (Figure 1), while Emma's work celebrates the positive progress made in Thailand (Figure 2). The images generated by Elena and Emma are replete with the rainbow-color scheme. Sandy, on the other hand, foregrounds the teacher identity of Jack by portraying him in professional settings with colleagues and students, with a limited hint of queer culture. In her photo essay, Sandy included a quote from Jack explaining how he approached talking about sexuality (Figure 3). Next to this paragraph, Sandy annotated that "I wanted to depict him as a <u>professor</u> [original emphasis], with other teachers."

Figure 2

*An excerpt from Emma's photo essay*

In a country celebrated as a global hub of queer culture, Bangkok, Thailand is often regarded as an oasis of acceptance for the LGBTQ+ community. It has gained a reputation for being a center for queer individuals seeking freedom and expression. But beneath the festivities and pride celebrations in Bangkok, Thailand does not have legal recognition of same-sex relationships. This legal gap leaves queer individuals without the legal protections and rights afforded to their heterosexual counterparts.

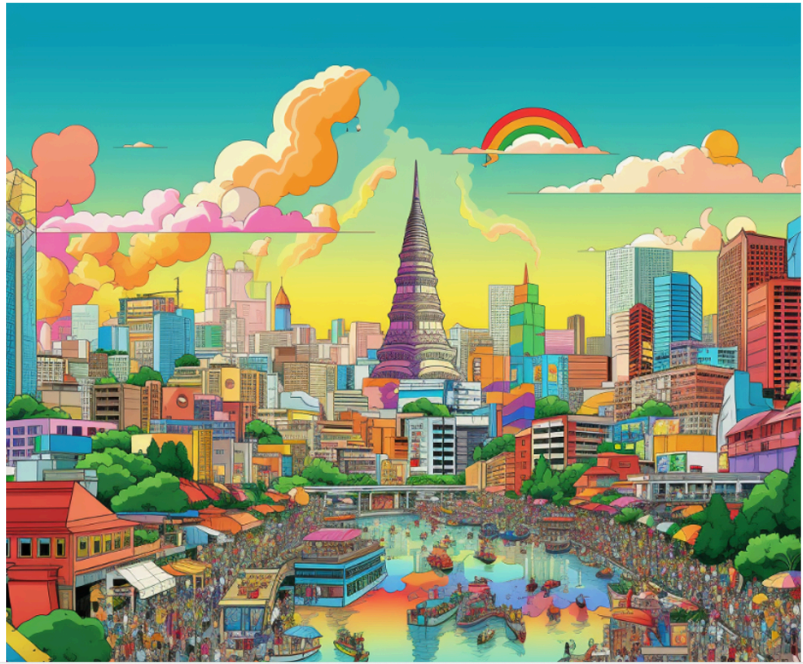

Figure 3

*An excerpt from Sandy's photo essay*

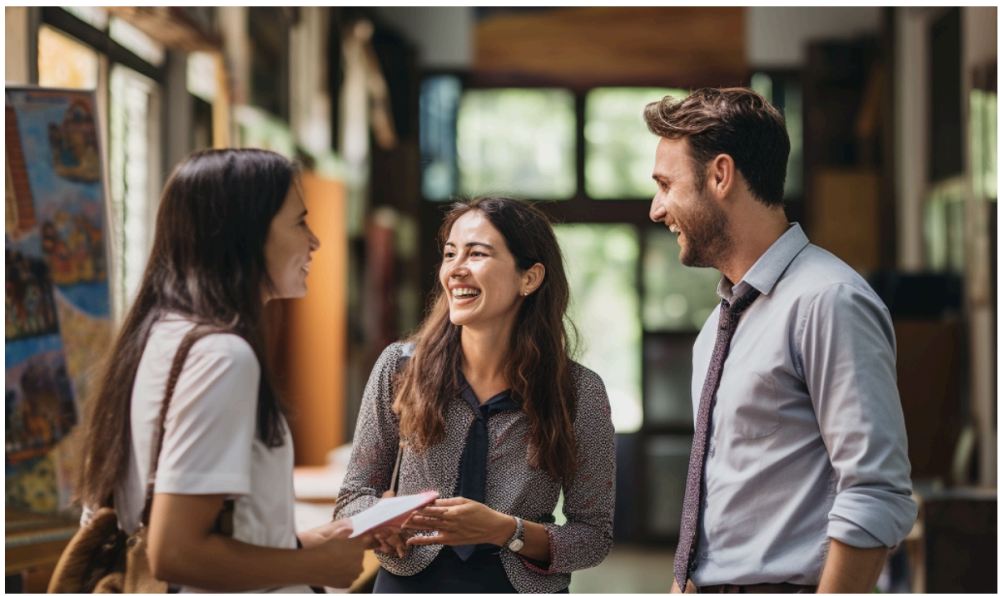

When asked about whether he would discuss his own sexuality in class, his response was:

"I don't talk about it [my sexuality] in class. I talk about it [sexuality] in general, like dealing with relevant discussions, such as skin colour, discrimination or sexuality. I don't try to avoid certain words. I just mention it as something normal... Perhaps it's uncommon that a dedicated teacher outs themselves in their class. I haven't, I think it's not common."

Moreover, students used AI-generated images to fulfill a wide variety of functions in relation to the text. The first function of images is to elicit emotions from readers. Harry, Linda, and Elena mentioned the power of visuals in appealing to emotions. When recounting Thuli's success in literacy development, Harry generated an image of a happily smiling lady in a classroom setting, with the accompanying quote from the main character Thuli that reads "Zulu, I'm perfect now. Only in English, now I'm struggling with English" (Figure 4). Harry explained his choice of words and images:

I kind of wanted to put that very proud smile on there. When you see a smile, you already know that she must be happy about something, right? ….So, that's a story right there, right? That I think invokes emotions.

Figure 4

*An excerpt from Harry's photo essay*

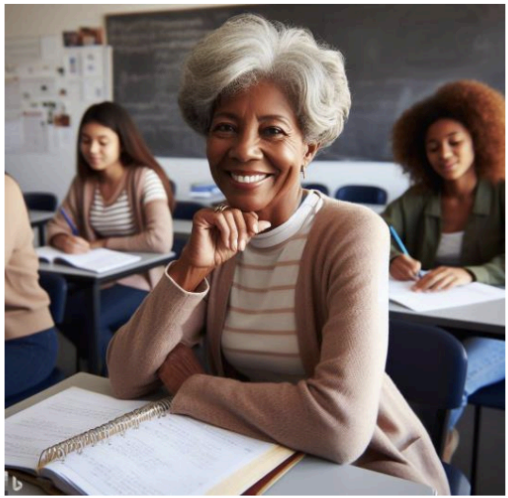

*"Zulu, I'm perfect now. Only in English, I'm struggling with English."*

In class, Thuli further enhances her literacy of English, smiling confidently (Bing)

The second commonly mentioned function of using visuals is to provide additional perspectives in storytelling. Examples in this category include using the change of photo texture to reflect the change of time (Nathen), depicting the evolution of identity with the change of brightness in images (Elena), and adding details and contextual information (Nathen, Linda, Emma, and Harry). For example, since the story of Sue unfolds chronologically, Nathen not only included images of Sue at different life stages but also deliberately prompted Midjourney to create images of higher quality toward the later stages. With the improved image quality, Nathen hops to show the "passage of time as the story progresses" (interview). More interestingly, Elena carefully attends to the mood of images, the change of which implies the subtle development of self-awareness experienced by the main character, as Elena explains: "they [Jack] start being kind of ominous, a little bit dark, a little bit sentimental, and then they move on to becoming a writer, perhaps they have more rainbows in them" (interview). Lastly, images are used to fill the gaps in storytelling by adding more details that are absent from the texts. For example, in an image of Sue teaching, Nathen selected the one with the biggest class size because he "associate(s) bigger class sizes with less funded schools," which aligns better with Sue's less privileged background.

Perhaps the most noteworthy function of using AI generated images is to concretize abstract ideas or metaphors into visual representations, which is identified in all participants' works. In some cases, students intentionally prompted AI systems to create such images. For example, Emma depicted Jack's personal success as him sitting "on top of the world" (Emma, interview) against the backdrop of a developing city (Figure 5). Two students attempted to visually represent the metaphor of "English as a master key to freedom" included in the original article. Ana portrayed this metaphor with a light shining through a half open door (Figure 6), while Harry illustrated the concept in a picture of Thuli holding a key in her hand (Figure 7). In both cases, the students pinpointed the metaphor in the captions. But at the same time, Harry also acknowledged that his image was fabricated in nature rather than reflecting real-world event:

> But the funny thing is that usually in the real world not many people hold a key with the English label on and smile at the camera, right? It's kind of a unique thing, but AI is able to generate whatever you want to imagine. So I think it helps us play with that imagination and that enriches our writer's voice.

Figure 5

*An excerpt from Emma's photo essay*

Jack's story shows the transformative influence of education in driving societal change. He captures how educators influence a student's academic growth but also their personal development. In a world advancing toward greater acceptance and inclusivity, Jack serves as a reminder of the challenges faced by queer individuals globally. He's dedicated to using his position as an educator to leave a positive impact on the world.

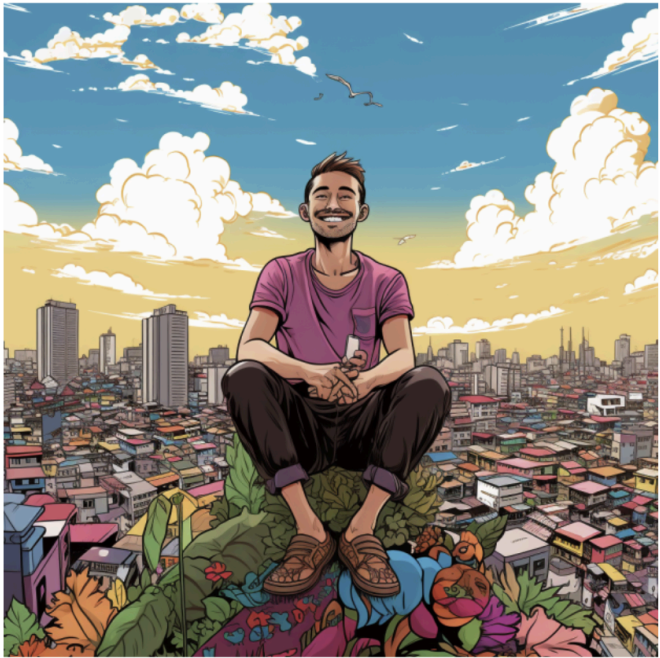

Figure 6

*The last image in Ana's photo essay*

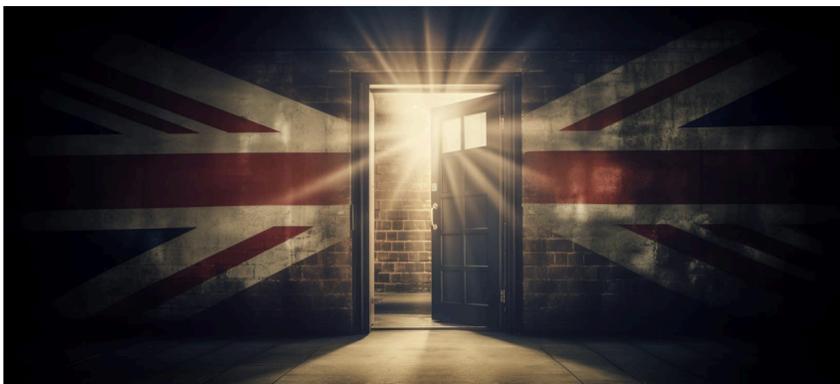

English as the light at the end of the tunnel, the "master key" unlocking countless doors, according to Thuli (Midjourney).

Figure 7

*The third image in Harry's photo essay*

*"It's like a master key, you've got a master key which opens each and every door. If you don't have English, you're just as good as a dead person."*

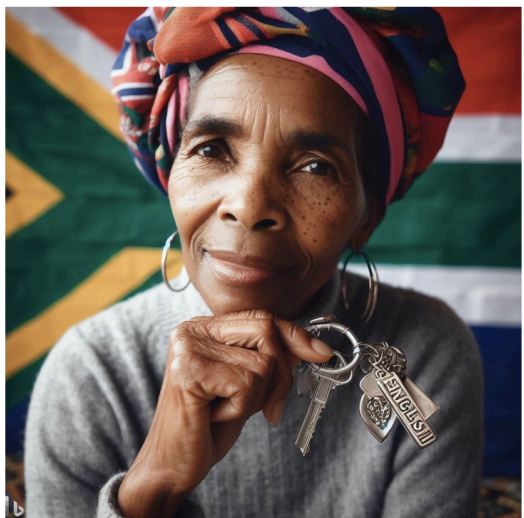

English is the master key for all doors (Bing)

Unlike the intentional use of images to concretize abstract ideas, Linda and Nathen noticed that AI systems tend to use symbols to represent concepts despite their lack of such commends. For example, in Linda's story, the protagonist Paul is an Englishman who lived in Germany. In an image depicting Paul's experience of living in the foreign country, Linda noted that Paul's nationality was implicitly symbolized in the blue-red color scheme of his scarf, which was not intended by the student herself. Similarly, Nathen wanted to generate an image showing the main character Sue's journey in Paris. Without explicit instruction to include Eiffel Tower, the AI system nonetheless generated an image of Sue standing in front of the landmark. Reflecting on this image, Nathen commented critically that: "I never mentioned that she [Sue] went to the Eiffel Tower. I don't know if that's true because she didn't talk about this in an interview, but it was just basically a stereotypical view of going to France" (interview).

In addition to using AI-generated images to achieve various rhetorical goals, students also articulated the functions of written texts in storytelling. The first function of written texts is to provide background and contextual information for storytelling. Harry and Emma started their photo essays with a paragraph describing the social and historical background, while Ana had a paragraph in the middle of her essay contextualizing her story, explaining that she needed words to "set up the background in the beginning" (interview).

Another unique affordance of written texts is to bring back the original voice of the main characters, which was done by incorporating direct quotations in narration. Three students (Elena, Nathan, and Sandy) discussed their reasons for using quotes from their protagonists. Elena explained that by including quotations from Jack, she "wanted to get a tone of how exactly he talked and how he felt about some things" (interview). Similarly, Nathen also tended to use direct quotations in places "where [he] couldn't gauge her [Sue's] emotions or outlook, or some of the really nuanced details" (interview). The attention to details is also reflected in Sandy's use of quotes: "I am writing about another person in a way that I can never capture exactly what they were thinking. The use of quotations is the most I can do to really capture what they were thinking" (interview).

**Blending.** Blending is the mix of genre features where "elements from different genres are combined to create a hybrid blend" (Hafner, 2015, p. 504). When approaching the unfamiliar genre of photo essay, students drew on their knowledge of other genres and incorporated genre features not restricted to photo essays. In terms of organizing the information, Ana attempted to structure her writing to resemble a "profile piece," which "start[s] with the individual story and then kind of generalize the experiences towards the later aspects" (Ana, interview). Nathen, on the other hand, structured his writing in a chronological order to reflect the features and layout of a biography.

Students also paid attention to the *register* of writing, intending it to be a mixture of both academic and public-facing writing. Nathen, for example, explained that he "tried to incorporate a good mix between being engaging and entertaining and then also including more general facts and her [Sue's] background" (interview). Similarly, Sandy wrote in the reflection that she "not only constructed an intimate, scholarly voice but also presented seriousness through photos." When asked to elaborate on her thoughts, Sandy said that "intimate" means establishing a closer author-reader relationship to capture the audience's attention. But at the same time, she also wanted her work to be "scholarly and academic" (Sandy, interview).

Lastly, while five out seven participants opted for more realistic depiction of the main characters and events, two students' works clearly deviated from that convention. Emma has included cartoon-style images throughout her photo essay, which was a result of both human agency and AI impact, as she discussed in the interview:

> As for the cartoon image, I think that was the first image that I generated sort of… it was already in that sort of cartoon style. And I really like the image. So I decided that I wanted to include him in a cartoon as well. I think it just added to the overall story, like the feel of this essay. And it was able to sort of include color and different dimensions that you probably wouldn't have gotten from a realistic image.

Another student, Harry, has chosen images of very different styles. While some of his images, such as the ones in Figure 4 and Figure 7, represent the realistic aspect of the story, others seem to capture complex feelings and emotions. As shown in Figure 8, the reaction toward linguistic oppression could be grotesquely depicted in the form of caricature, as Harry explained: "I was trying to represent the anguish, the burning books, the burning literacy, the burning out alphabets, burning freedom, basically" (interview).

Figure 8

*An excerpt from Harry's photo essay*

One of the methods that the British used to control and contain the natives of South Africa during colonial rule was the *Bantu Education Act of 1953* that forbid black learners at non-registered schools and forced Afrikaans as the only language to be taught and learned. Afrikaans is the language of the colonizers, the oppressors, the enslavers from the Dutch-Portuguese creole. The British government administered a rule on language; by controlling the communication and putting a limitation, the colonization further seized freedom.

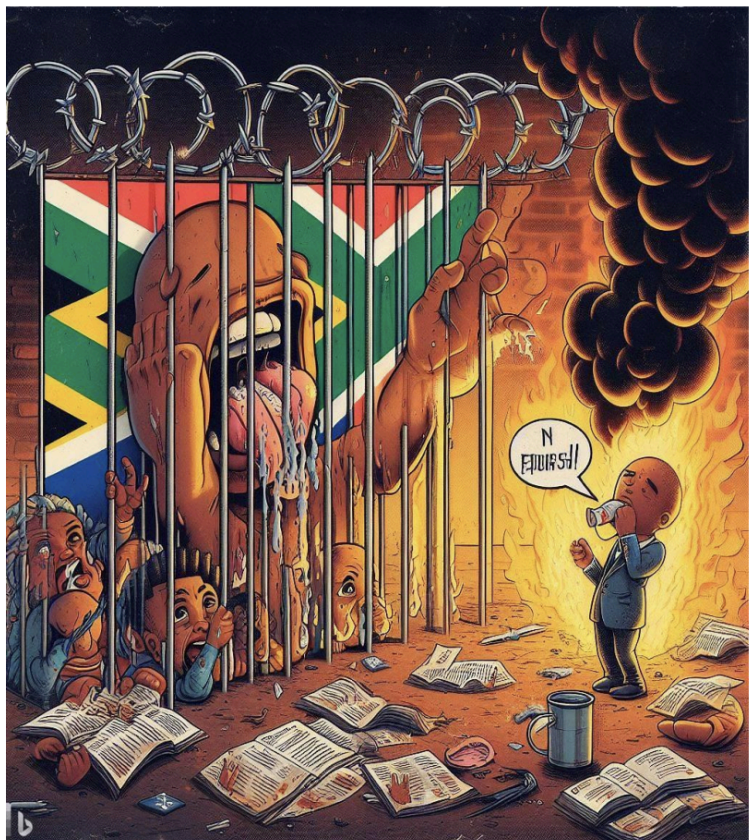

Suppression of Language in South Africa (Bing)

*It used language as a weapon to stifle.*

This section presents the sophisticated and deliberate decision-makings that students engaged in during multimodal writing. Informed by their understanding of voice, students carefully considered how available resources could be remixed to accurately and creatively express their authorly intent, evoke strong emotional connections with the audience, and highlight unique perspectives of the remediated story.

### *4.2 The Affordances of Using AI in Multimodal Writing*

Students generally perceived four aspects of the affordances of using AI-powered image generating tools in composing a photo essay: high quality of the images, pleasant composing experience, satisfactory composing outcome, and facilitated voice expression. The first affordance is concerned with the quality of AI-generated images, as they were depicted as aesthetic, realistic, creative, and in-depth. Specifically, some described the pictures from Midjourney as "aesthetically pleasing and realistic" (Ana, interview) and opening a creative outlet for the author's creativity as it gives authors an idea of what they want, which normal pictures would not be able to do (Sandy, interview). Others commented that the images produced by AI could pick up on nuanced details (Nathen, interview) and usually were generated rather quickly (Emma, interview). In addition, Sandy elaborated on how AI-generated images satisfied her needs:

> Generally I like the tone that AI gave me. I was pretty happy with [it]. I didn't want it to be too dark of a picture or like black and white or anything like that. And then they were all pretty vibrant with colors. And the style of the pictures were pretty consistent (Sandy, interview)

The second affordance perceived by the students is the pleasant composing experience enhanced by AI. For example, Linda described the experience of writing a photo essay with the AI-powered tool as "interesting and intriguing" (interview) and believed that Midjourney offered more flexibility for her to select the image for use among a wide array of options. Nathen likewise acknowledged the flexibility offered by AI, explaining that Midjourney allowed him to manipulate different layers of expression in a virtual world by switching between realistic and fake pictures. Harry, on the other hand, projected his sight further, perceiving the use of AI in composing a photo essay as an opportunity and experience for him to mull over what the future of AI technology can offer.

The next prominent affordance, in students' perspectives, is the positive composing outcome that AI may amplify. Harry pointed out that assuming what AI produced was accurate, it may enrich the audience's impression on the topic as the readers' understanding can be increased on the given topic by looking at AI-generated images in the condition that the author does not have first-hand experience about the written topic. The last affordance is related to authors' voice expression. A few students mentioned that AI-generated pictures add to the uniqueness of the writer's work (Linda), make the author's voice more visible and concrete (Emma), and help with emotion expressions (Elena and Harry). For example, as shown in the previous section (see Figure 4), Harry placed emphasis on the emotional aspects of storytelling and purposely included images that appeal to pathos. These examples have highlighted the flexibility and creativity that students are allowed when approaching multimodal writing with the assistance of AI tools.

### 4.3 Challenges of Constructing Voices

The participants encountered various challenges while utilizing AI-powered image generation tools to express their voice. One major challenge highlighted by students was the difficulty in effectively presenting and translating their ideas into visual formats. Specifically, they encountered three hurdles during the process: (1) visualizing a narrative or an abstract idea, (2) articulating an envisioned scene, and (3) lacking sufficient details for visualization. Several participants reported struggling with converting complex stories and abstract concepts into tangible components of visual images. As Ana stated, "It's hard when I have a narrative, an idea, or part of a narrative and want to make it something that can be presented visually" (interview).

The students wanted to identify concrete objects and elements that could represent an abstract idea or intricate components in a narrative, so the AI tool could operationalize their vision. Ana further elaborated, "It was really hard to figure out what components should be in the photo. Like how do I represent English? How do I represent an abstract idea into those photos?... I can't visualize the education act itself. I also can't visualize English itself without finding a symbol of English" (interview). Similarly, Nathen admitted that he could only envision broad scenes that lacked specificity. He explained, "Mine would be generally like the perspective, the colorization, like how Sue would look in the photo, because I was providing a pretty general description about her" (interview). These students' difficulty stems from the challenge of converting ideas into concrete visual elements that AI can comprehend and materialize.

For others, the obstacle lies in the translation of mental images into descriptive language that AI tools can process. As Linda pointed out in the interview, "I found that it was quite challenging, personally, to use Midjourney a little bit, because I don't consider myself the most creative person when it comes to describing a picture thing." The third obstacle identified by the students was the limited information available in the original text, which to them, was essential for detailed and authentic representations of the text information. The students sought to get richer details so they could accurately convey the intended emotions, objects, and meanings. As Elena mentioned,

> It was hard to visualize the person in the essay, because the essay itself does not tell us a lot about the person other than the main part of his identity, which is the fact that he's queer… So when I'm giving AI a prompt, I can't say where he's from, or what his age is, or anything like that.

The second major challenge students experienced seems to result from the inability of the AI system in accurately producing desirable outcomes, characterized by the following four specific gaps: (1) irrelevance, (2) lack of specificity, (3) inaccuracy, and (4) presumptive interpretations. Emma, for instance, noted that AI sometimes produced images unrelated to her intended message, saying that "It would give me images that I felt like just didn't relate to what I was trying. They didn't convey the image that I wanted to convey." Elena echoed this statement, pointing out AI's inability to get "exactly what I am thinking of." She also observed that the AI might fail to capture the precise elements she wanted: "It just cannot get the image that was in my head as specific as I am." However, while unexpected, Elena viewed this gap "more as a boon than vain" as she believed that it could facilitate her creative thinking by "getting new images."

Other students also acknowledged that AI might twist the original meaning and cannot give them the most accurate image they wanted. Linda recounted an instance where the AI's outputs might constrain her writer's voice as they provided inaccurate images that missed her original intentions. As she explained,

> For this first image, I kind of wanted the people around the table looking a little bit more upset or confused, but I could only get them kind of also looking excited and rowdy. I mean, they do show some concern. But I think that AI kind of yielded the full image that I was expecting…(Linda, interview)

Another important challenge that the students had was working with AI's assumptions or possible prejudice based on the information they put in. For instance, Sandy, when prompting AI to generate an image depicting students sitting around a projector, AI created an image (see Figure 9) in which students all sat on the ground even though she did not ask for this seating arrangement. Sandy speculated that AI made biased assumptions about Thai students, as she reasoned

> Because I said Thai students, I guess that's where the background came from. Perhaps AI assumes that Thai students wouldn't be as rich to have a fancy movie theater, like a fancy projector. So that's what came out. It has prejudice, I guess. (Sandy, interview)

> Echoing Sandy, Nathen commented that AI, much like humans, relied on the provided and often incomplete information to make assumptions, and this might lead to a recursive cycle of misinterpretation, further complicating his pursuit of authentic visual representation. As he mentioned, "Midjourney was doing the same exact thing that I had done, which was to draw on the information I provided to it and then make assumptions to generate an even more holistic product."

Figure 9

*AI depiction of Thai students sitting around a projector in Sandy's photo essay*

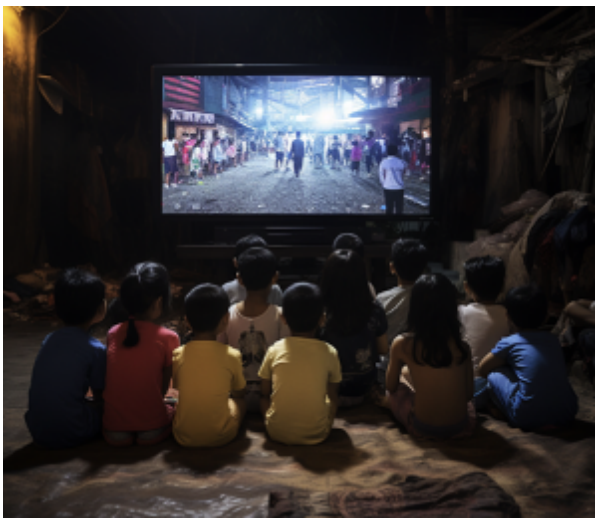

The third major challenge, as experienced by many students, involves the difficulty of ensuring AI-generated images maintain consistency in both style and the depiction of objects. Although Sandy was satisfied with the consistent style of AI-generated images, some students reported otherwise. Emma, for example, noted her attempts to preserve a uniform style across images. She mentioned, "Between all of the images, I tried to maintain the same style. For example, like the cartoon style, but it was kind of difficult for AI to generate that same exact style of cartoon every time, because AI doesn't know exactly what you want in your mind." Similarly, Sandy and Linda encountered difficulties in "generating the same person" more than one time with Midjourney. As can be seen from the students' experiences, achieving a consistent voice through AI-generated images is complicated by the inherent unpredictability of AI tools.

Other challenges that students experienced also include engineering the prompts, adapting to the learning curves of using AI image generation tools, and dealing with "knowledge comprehension gaps" (Li et al., 2023). Harry mentioned that learning how to write effective prompts was "the first challenge." Other participants also recognized that they experimented with various prompts and phrasings to inch closer to their envisioned results. Students also acknowledged the learning curves of adapting to the limitations of AI. Ana, for instance, was "surprised" to discover AI's inadequate grasp of complex concepts like apartheid. Likewise, Harry recognized that the process of understanding "how the algorithm works" is a process of "trial and error." Additionally, students might also experience a "knowledge comprehension gap" (see Li et al., 2023), referring to the gap in fully understanding AI's outputs, as illustrated by Harry:

> The first picture here [See Figure 10] I wanted to put the flag of South Africa because that's where the lead quest starts from. And there are different great people in history who have done great deeds. Unfortunately, I don't know what every single person did in that picture...I keep looking at all the people in there and I think I searched for a few of them, but I wish that AI generated a picture that could tell me who those people are actually in the collage in the beginning of the picture, there's so many people there.

Figure 10

*Harry's image on South Africa's history generated by Bing*

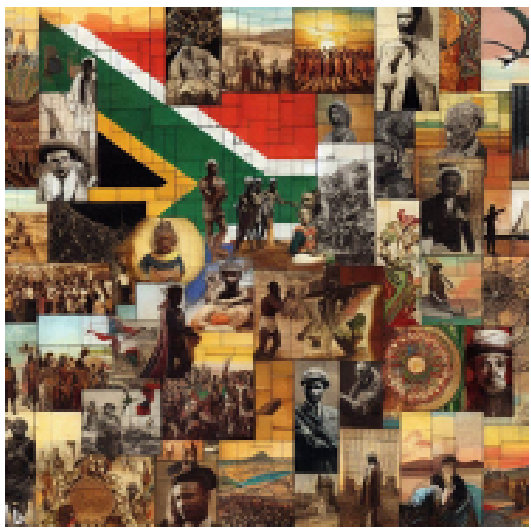

## 5. Discussion

The study investigates how AI image generating tools could be leveraged in multimodal writing to enhance students' voice construction. As the findings suggest, students have developed sophisticated understandings about the concept of voice, placing slightly different emphasis on the author, audience, and the interaction of the two (Tardy, 2012). Unlike the students in Hafner's study (2015) who had to choose from online image banks, students in our study enjoyed more flexibility in designing their multimodal works, drawing on the affordances of various modes to create engaging stories. In constructing a designer's voice in multimodal writing, students paid special attention to maintaining consistency across modes, as well as within the same mode. This was achieved by strategically prompting AI tools to generate images that align with their rhetorical purposes and their understanding of the genre. As illustrated by multiple data, when remixing various resources, students carefully negotiated the rhetorical situation, modal affordances, and genre conventions. As a result, students have created photo essays that appear drastically different from each other even when remediating the same story, such as the works of Elena, Emma, and Sandy. These findings highlight the rhetorical nature of remixing in multimodal writing, shed light on the affordances of using AI tools to enhance one's rhetorical intention, and underline the pedagogical values of engaging students in remixing activities (Edwards, 2016; Dusenberry et al., 2015; Hafner, 2015; Palmeri, 2012).

Adding AI image generating tools to the remixing process presents unique opportunities for voice expression and challenges. One of the biggest opportunities, as shown in this study, is to engage students in a recursive, trial-and-error learning process that is fun and motivating for most learners. To fully express their voice without compromise, students need to not only visualize an idea mentally but also translate that mental image into precise, descriptive language. This process requires students to learn how to prompt and navigate AI tools to produce desired outputs, critically evaluate the generated results, and re-work with AI to fine-tune these outcomes to ensure a strong alignment between the student writers' objectives and AI's interpretative algorithms. The process is not linear but iterative, demanding students to continuously adjust their approach based on the feedback loop between their creative vision and the AI's rendered outcomes (Henrickson & Meroño-Peñuela, 2023). The iteration opens up opportunities for students to learn not only about the functionalities, potentials, and constraints of AI tools but also about their own creative expression (Author, year).

However, to successfully express one's voice with the assistance of AI is not an easy task. As reiterated in the interviews, students sometimes find it hard to effectively present and translate their ideas into visuals. This could partly be attributed to students' insufficient prompt literacy, which is defined by Kang and Li (2023) as "the trained ability or knowledge to appropriately and effectively formulate and adjust prompts, including visual or auditory stimuli, based on specific contexts and purposes" (p. 3). The best practice to scaffold prompt literacy development, as Warschauer et al. (2023) suggest, is through explicit and sustained guidance on how to construct and refine prompts. Another possible reason for the perceived difficulty has to do with how students understand the relationships among different modes (Kang & Li, 2023). As suggested by the findings, many students felt it necessary to reiterate an abstract idea or metaphor discussed in the original article in visuals, thus pursuing a one-on-one image-text relationship. In fact, images and texts exist in complex relationships (Marsh & White, 2003), and it is rare to see abstract ideas being concretized in photo essays (e.g., the metaphor of "English as a key" being represented in an actual key). While the current study was not specifically designed to help students develop prompt and multimodal literacy, the findings suggest room for adapting the photo essay assignment to achieve these goals in the future.

At the same time, we should also note that AI image-generating tools are inherently limited and biased (Aguillar, 2024), which further contributes to the difficulties of using it in multimodal writing. One of the biggest limitations is the unpredictable nature of AI-generated outputs (Bryd et al., 2023). This presents a challenge for students to maintain the same style throughout their work despite their best intentions. Another limitation of AI systems is the tendency to reinforce stereotypes and even biases. As Nathen and Linda reported, the AI system tends to represent abstract ideas, such as nationality, in widely recognizable symbols. While this is an easy way to get the message across to the readers, doing so also reduces the complexity of these ideas. The AI-created images, when used uncritically, could strengthen the stereotypes already associated with certain cultures and communities. Even more alarming are the biases that AI tools demonstrate toward less developed nations, as observed by Sandy in her AI-generated image of Thai classrooms. As highlighted in recent publications on AI in composition studies (Aguillar, 2024; Owusu-Ansah, 2023), AI technology is never neutral because of the innate biases built into the design, and such biases need to be addressed explicitly in class discussions to help students become more responsible and ethical users.

In terms of the assignment design, we would like to highlight a few points that deserve further consideration. The first is the task appropriateness of incorporating AI tools. As Ana explained, students who create fictional genres might benefit more from the hyper-real and creative images generated by AI systems. In fact, recent publications have demonstrated the successful applications of AI in creative writing (e.g., Easter, 2023). Using AI tools in historical storytelling, however, demands more scrutiny on the issues of authenticity, as such images could create a sense of detachment from the historical events. Moreover, the labor-intensive nature of interacting with and learning to use AI tools, as many participants have experienced, should also be taken into consideration when teachers employ AI tools as a teaching resource.

The study is limited in the following ways. First, the exploratory nature of this study places a constraint on how many participants we were able to recruit. The small number of participants, thus the limited data size, calls for caution when interpreting the findings. Second, students' writing processes are analyzed using their interview data, triangulated by their written reflection, annotations, and photo essay products. While the current data collection helps us understand what students considered and paid attention to when attempting to complete this task, it does not capture all the actions taken during the process, especially students' input into the AI systems. Future studies could delve deeper into students' multimodal writing with AI by documenting and analyzing their prompting process as an integral part of this activity.

## 6. Conclusion

This study investigates how students mobilize semiotic resources and AI tools to construct voice in multimodal writing. The findings have shed light on students' purposeful remixing practices as afforded by AI image-generating tools, while also revealing the challenges students encountered during the process. This study not only contributes to the burgeoning literature of teaching composition with AI, but also provides empirical data for understanding the concept of voice in multimodal context. The current study focuses primarily on student authors' choices, decision-making, and negotiation when constructing voices. Future studies could adopt a dialogical approach to survey how readers actively construct the author's voice in multimodal writing. Second, as discussed earlier, this assignment presents valuable opportunities for students to grapple with the new genre, develop prompt and multimodal literacy, and critically reflect on the ethical issues involved in using AI. We need to continue our effort of exploring the effective ways of incorporating AI in composition pedagogy, while also examining how students are cognitively, behaviorally, and emotionally engaged in using different AI tools in multimodal writing.